\documentclass[10pt]{iopart}
\usepackage{epsfig,cite}
\usepackage{latexsym}

\newcommand{\beq}{\begin{eqnarray}}
\newcommand{\eeq}{\end{eqnarray}}
\newcommand{\be}{\begin{eqnarray*}}
\newcommand{\ee}{\end{eqnarray*}}

\newcommand{\QQ}{\scriptscriptstyle{Q \bar{Q}}}
\newcommand{\cc}{{c\bar{c}}}

\begin{document}

\title{Resolving the $J/\psi$ RHIC puzzles at LHC}

\author{
Bravina~L~V\dag, Tywoniuk~K\ddag, Capella~A\S, Ferreiro~E~G\ddag,
Kaidalov~A~B$\Vert$, Zabrodin~E~E\dag\P
}
\address{\dag\
         Department of Physics, University of Oslo, PB 1048 Blindern,
         N-0316 Oslo, Norway}
\address{\ddag\
         Departamento de F{\'\i}sica de Part{\'\i}culas, Universidad de 
         Santiago de Compostela, 15782 Santiago de Compostela, Spain}
\address{\S\
         Laboratoire de Physique Th\'eorique, Universit\'e de Paris XI, 
         B\^atiment 210, 91405 Orsay Cedex, France}
\address{$\Vert$\
         Institute of Theoretical and Experimental Physics, RU-117259 
         Moscow, Russia}
\address{\P\
         Institute for Nuclear Physics, Moscow State University,
         RU-119899 Moscow, Russia}

\begin{abstract}
Experiments with gold-gold collisions at RHIC have revealed (i) 
stronger suppression of charmonium production at forward rapidity 
than at midrapidity and (ii) the similarity between the suppression 
degrees at RHIC and SPS energies. To describe these findings we
employ the model that includes nuclear shadowing effects, calculated
within the Glauber-Gribov theory, rapidity-dependent absorptive 
mechanism, caused by energy-momentum conservation, and dissociation 
and recombination of the charmonium due to interaction with co-moving
matter. The free parameters of the model are tuned and fixed by 
comparison with experimental data at lower energies. A good agreement
with the RHIC results concerning the rapidity and centrality 
distributions is obtained for both heavy {\it Au+Au} and light 
{\it Cu+Cu} colliding system. For {\it pA} and {\it A+A} collisions 
at LHC the model predicts stronger suppression of the charmonium and 
bottomonium yields in stark contrast to thermal model predictions. 
\end{abstract}


\section{Introduction}
\label{intro}

The investigation of nuclear matter under extreme temperatures and
densities, and the search for a predicted transition to a deconfined
phase of quarks and gluons, the so-called Quark-Gluon Plasma (QGP),
is one of the main goals of heavy-ion experiments at ultrarelativistic
energies. Both theorists and experimentalists are looking for genuine
QGP fingerprints, that cannot be masked or washed out by processes on
a hadronic level. Charmonium was proposed about two decades ago
\cite{SM86} as one of the most promising QGP messengers because its
yield would be significantly suppressed due to color Debye screening in 
the plasma phase. Also, due to the small interaction cross section of 
$J/\psi$ in hadronic matter, charmonium spectrum is expected to carry 
information about the early hot and dense stage of nuclear collision.
Since the volume of the produced QGP depends on the collision energy, 
centrality and mass of colliding nuclei, it is generally believed that 
the suppression of the $J/\psi$ yield would increase with rise of the 
aforementioned factors.

Therefore, the PHENIX measurement of the nuclear modification factor of
charmonium at top RHIC energy $\sqrt{s} = 200$\,AGeV \cite{Ad07} 
uncovered at least two unexpected features. Firstly, compared to
charmonium suppression in lead-lead collisions at SPS energy 
$E_{lab} = 160$\,AGeV \cite{Al05} the level of suppression at 
midrapidity at RHIC was found to be quite similar for the same number
of participants despite the order of magnitude difference in 
center-of-mass energies of heavy-ion collisions. Secondly, $J/\psi$ 
suppression increases unambiguously with rising rapidity, whereas the
highest energy density and the most dense medium should be produced
at $y = 0$. These findings, as well as
the experimental results for $d Au$ collisions, where the plasma 
formation is very unlikely, attract the attention to the whole
variety of the processes, including both initial and final state 
effects, that are responsible for the charmonium production and its
propagation through hot and dense medium. 

The paper is organized as follows. Section~\ref{model} presents a
description of the model that contains a comprehensive treatment of 
initial-state nuclear effects, such as nuclear shadowing and nuclear 
absorption, and final state interactions with the co-moving matter.
Comparison with the available experimental data and predictions for
$Pb+Pb$ collisions at LHC energy is given in Sec.~\ref{resul}. Finally, 
conclusions are drawn in Sec.~\ref{concl}.

\section{Description of the model}
\label{model}

Nuclear effects in nucleus-nucleus collisions are usually expressed through
the so-called nuclear modification factor, $R^{J/\psi}_{AB}
(b)$, defined as the ratio of the $J/\psi$ yield in {\it A+A} and
{\it pp} scaled by the
number of binary nucleon-nucleon collisions, $n(b)$. We have then
\beq \label{eq:ratioJpsi}
R^{J/\psi}_{AB}(b) \;&=&\;
\frac{\mbox{d}N^{J/\psi}_{AB}/\mbox{d}y}{n(b)
  \,\mbox{d}N^{J/\psi}_{pp}/\mbox{d}y} \nonumber \\ 
\;&=&\; \frac{\int\mbox{d}^2s \, 
  \sigma_{AB}(b) \, n(b,s) \, S_{J/\psi}^{sh}(b,s) \, 
  S^{abs}(b,s) \, S^{co}(b,s)
}{\int \mbox{d}^2 s \, \sigma_{AB} (b) \, n(b,s)} \;.
\eeq
Here $\sigma_{AB}(b) = 1 - \exp [-\sigma_{pp}\, AB\, T_{AB}(b)]$, 
$T_{AB}(b) = \int\mbox{d}^2s T_A(s)T_B(b-s)$ is the nuclear overlapping
function, $T_A(b)$ is obtained from Woods-Saxon nuclear densities, and 
\beq
\label{eq:nbin}
n(b,s) \;=\; \sigma_{pp} AB \, T_A(s)\, T_B(b-s)/\sigma_{AB}(b)\;,
\eeq
where upon integration over $\mbox{d}^2s$ we obtain the number of
binary nucleon-nucleon collisions at impact parameter $b$, $n(b)$.
The three additional factors in the numerator of
Eq.~(\ref{eq:ratioJpsi}), $S^{sh}$, $S^{abs}$ and $S^{co}$, denote the
effects of shadowing, nuclear absorption, and interaction with the 
co-moving matter, respectively. Let us discuss them briefly.

The nuclear absorption is usually interpreted as suppression of 
$J/\psi$ yield because of multiple scattering of a $c\bar{c}$ pair 
within the nuclear medium.
At low energies the primordial spectrum of particles created in
scattering off a nucleus is mainly altered by {\bf (i)} interactions 
with the nuclear matter they traverse on the way out to the detector 
and {\bf (ii)} energy-momentum conservation. For {\it A+A} collisions 
these effects can be combined into the generalized suppression factor
\beq
\label{eq:Sabs}
S^{abs} &=& \frac{
        \left[1 - \exp(-\xi(x_+) \sigma_{\QQ} AT_A(b)) \right]}
       {\xi(x_+)\xi(x_-) \sigma_{\QQ}^2 AB \, T_A(s) T_B(b-s)} 
\nonumber \\
&\times& \left[1 - \exp (-\xi(x_-) \sigma_{\QQ} B T_B(b-s)) \right]\;,
\eeq
where $x_\pm = (\sqrt{x_{\rm F}^2 - 4M^2/s} \pm x_{\rm F})/2$, and 
$\xi(x_\pm) = (1-\epsilon) + \epsilon x_\pm^\gamma$ determines 
both absorption and energy-momentum conservation.
In \cite{Bor93} it has been found that $\gamma=2$, $\epsilon=0.75$ 
and $\sigma_{\QQ} = 20$ mb give a good description of data. This
corresponds to $\sigma_{abs} = 5$ mb at mid-rapidity and agrees well
with other studies. 

Secondly, coherence effects will lead to nuclear shadowing for both soft
and hard processes at RHIC, and therefore for the production of
heavy flavor. Shadowing can be calculated
within the Glauber-Gribov theory \cite{Gr69}, and we will utilize
the generalized Schwimmer model of multiple scattering
\cite{Sch75}. In this case the second suppression factor in
Eq.~(\ref{eq:ratioJpsi}) reads
\beq
\label{eq:schwimmer}
S^{sh}(b,s,y) \;=\; \frac{1}{1 \,+\, A F(y_A) T_A(s)} \, \frac{1}{1
  \,+\, B F(y_B) T_B (b-s)} \;,
\eeq
where the main contribution to the function $F(y)$, that encodes the 
dynamics of shadowing, comes from the gluon rather than from the quark
shadowing \cite{plb07,plb08}.
At SPS energy the nuclear absorption dominates over the shadowing,
whereas RHIC already belongs to the high-energy
regime. Nuclear shadowing is non-negligible at mid-rapidity, and the
combined effect of shadowing and energy-momentum conservation should 
be accounted for at forward rapidities. At LHC shadowing will be 
very strong even at $y=0$, while energy-momentum conservation becomes 
a minor effect.

Finally, the processes of dissociation and recombination of $c\bar{c}$
pairs in the dense medium should be taken into account. We employ the 
co-movers interaction model (CIM) \cite{Cap05} that was recently
modified to incorporate the recombination mechanism into consideration
\cite{epjc08}.
Assuming a pure longitudinal expansion and boost invariance of the 
system, the rate equation which includes both dissociation and 
recombination effects for the density of charmonium at a given 
production point at impact parameter $s$
reads
\beq
\label{eq:rateeq}
\tau \frac{d N_{J/\psi} (b,s,y)}{d \tau} &=& -\sigma_{co} 
\Big[ N^{co}(b,s,y) N_{J/\psi}(b,s,y) \\
   & & - \, N_c (b,s,y) N_{\bar{c}}(b,s,y) \Big] \;, \nonumber
\eeq
where $N^{co}$, $N_{J/\psi}$ and $N_{c (\bar{c})}$ is the density of 
comovers, $J/\psi$ and open charm, respectively, and $\sigma_{co}$ is 
the interaction cross section for both dissociation of charmonium with 
co-movers and regeneration of $J/\psi$ from $\cc$ pairs in the system
averaged over the momentum distribution of the participants. It is the 
constant of proportionality for both the dissociation and recombination 
terms due to detailed balance $ N_{J/\psi}(b,s,y)\, N^{co}(b,s,y) = 
N_{c}(b,s,y)\,N_{\bar{c}}(b,s,y).$
The solution of Eq.~(\ref{eq:rateeq}) can be approximated by
\beq
S^{co}(b,s,y) &=& \exp \Big\{ - \sigma_{co} \big[ N^{co}(b,s,y) 
\nonumber \\
  &-& C(y) N_{bin}(b,s) S^{shad}(b,s,y) \big] \, \ln \big[
  \frac{N^{co}}{N_{pp}(0)} \big] \Big\} \;,\nonumber
\eeq
where
\beq
\label{eq:C}
C(y) \;=\; \frac{\left( d \sigma^\cc_{pp} \big/ dy \right)^2}
{\sigma^{ND}_{pp} d \sigma^{J/\psi}_{pp} \big/ dy} \;.
\eeq
Details of the model can be found in \cite{epjc08}. The 
quantities in Eq.~(\ref{eq:C}) are all related to $pp$ collisions at 
the corresponding energy and are taken from experiment. Note, that
the extension of the CIM to the recombination effects does not imply 
any additional parameters.

\section{Heavy quarkonium at RHIC and LHC}
\label{resul}

\begin{figure}[t!]
  \begin{center}
    \includegraphics[width=.5\linewidth]{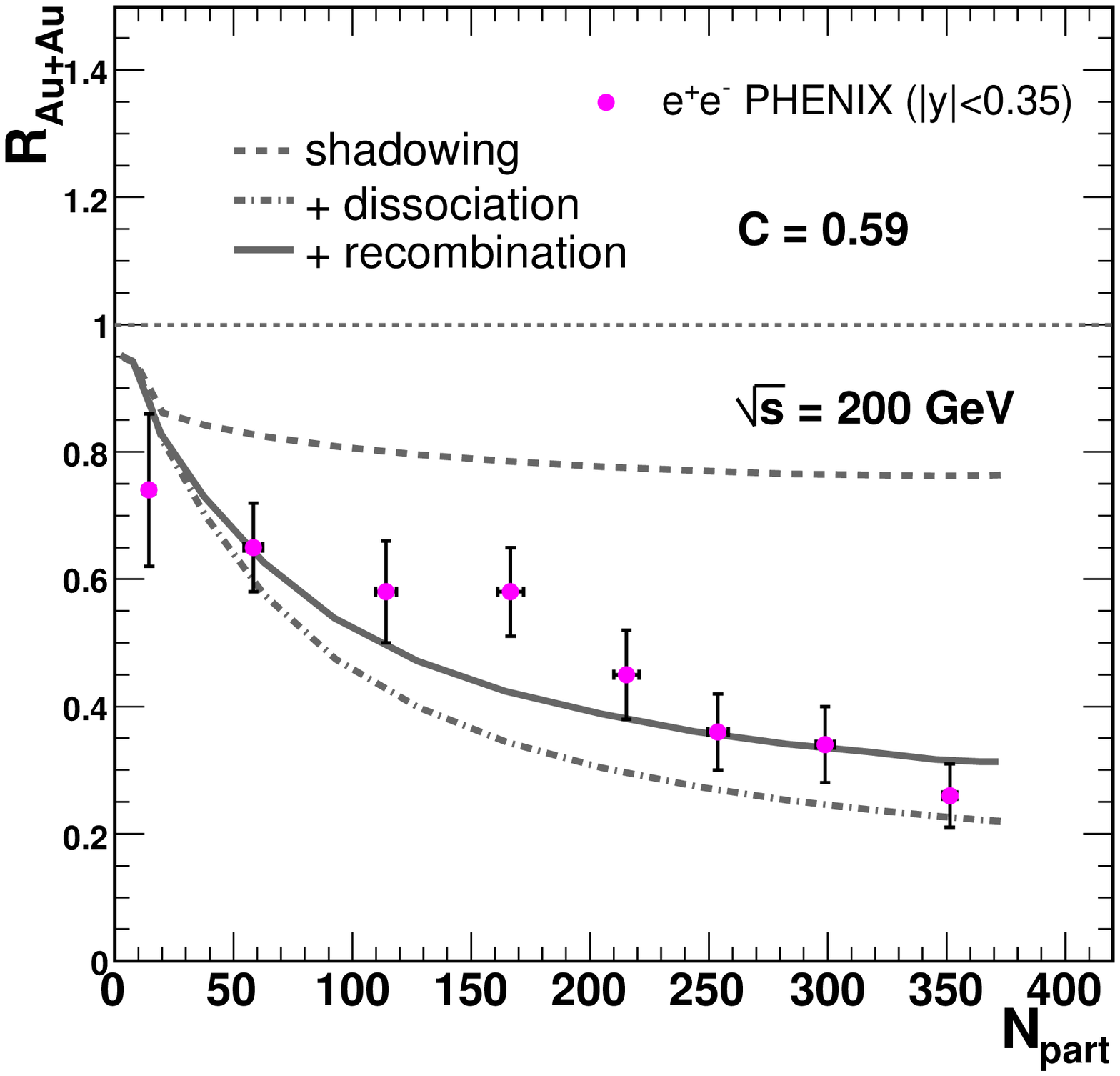}%
    \includegraphics[width=.5\linewidth]{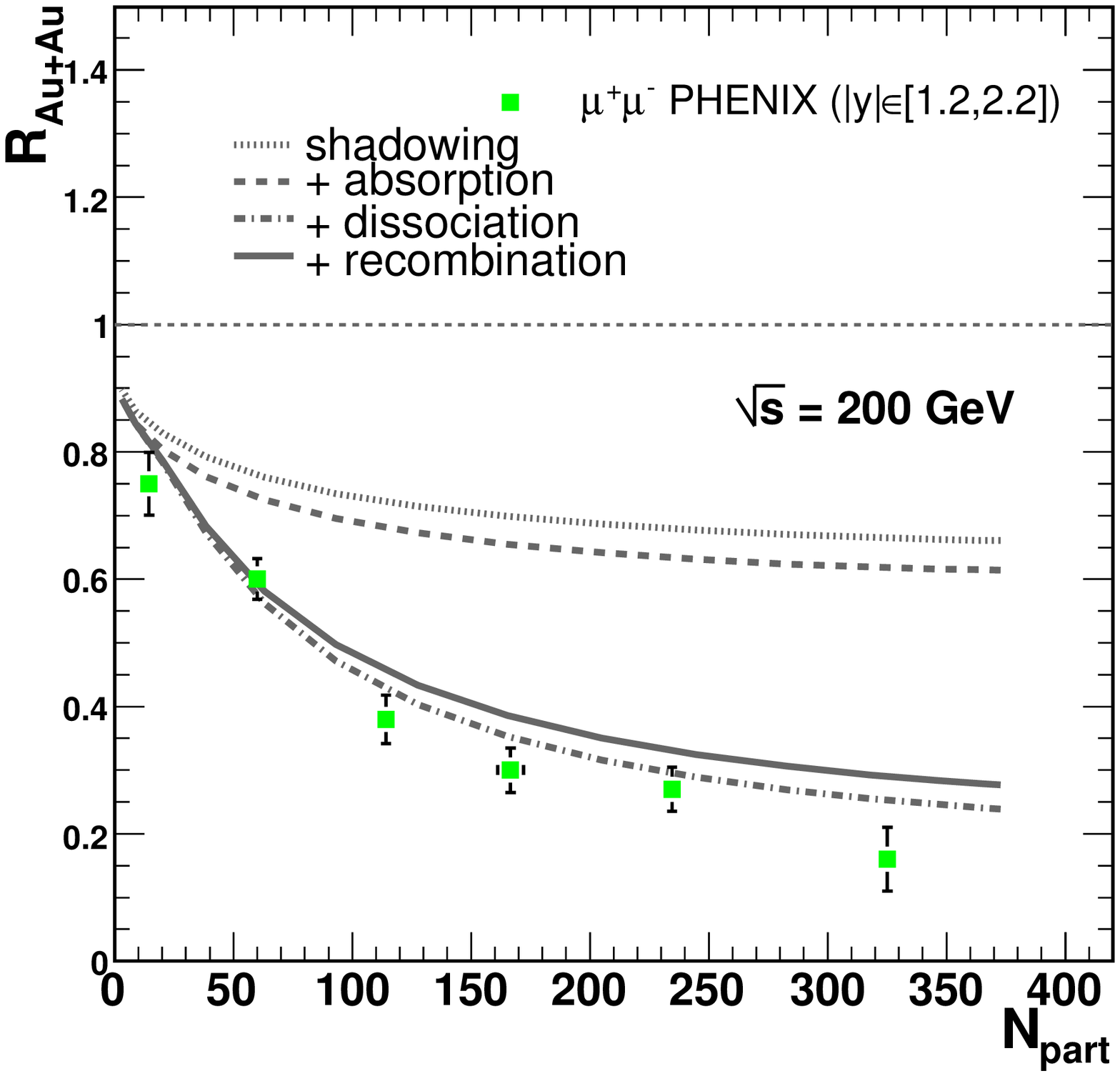}
  \end{center}
  \caption{Results for $J/\psi$ suppression in {\it Au+Au} at RHIC
    ($\sqrt{s} = 200$ GeV) at mid- (left 
    figure), and at forward rapidities
    (right figure). Data are from \cite{Ad07}. The solid curves
    are the final results. The dash-dotted ones are the results
    without recombination ($C = 0$). The dashed line is the total
    initial-state effect. The dotted line in the right figure is the
    result of shadowing. In the left figure the last two lines
    coincide.}
  \label{fig1}
\end{figure}

The density of open and hidden charm at mid-rapidity in {\it pp} 
collisions at $\sqrt{s} = 200$ GeV has been reported in \cite{Ad06}.
In the left picture of Fig.~\ref{fig1} we present the results of our
model compared to experimental data at mid-rapidity. The different
contributions to $J/\psi$ suppression are shown.
Note that at mid-rapidities the initial-state effect is just the
shadowing. As discussed above nuclear absorption due to
energy-momentum conservation is present at forward rapidities but is
negligibly small at mid-rapidities. One can see that the $J/\psi$ 
suppression at forward rapidity is somewhat larger than that at 
mid-rapidities, in full accord with experimental data. This is due 
to both the recombination term and the initial-state effects. The 
latter are stronger for forward rapidities. 

\begin{figure}[t!]
  \begin{center}
    \includegraphics[width=.5\linewidth]{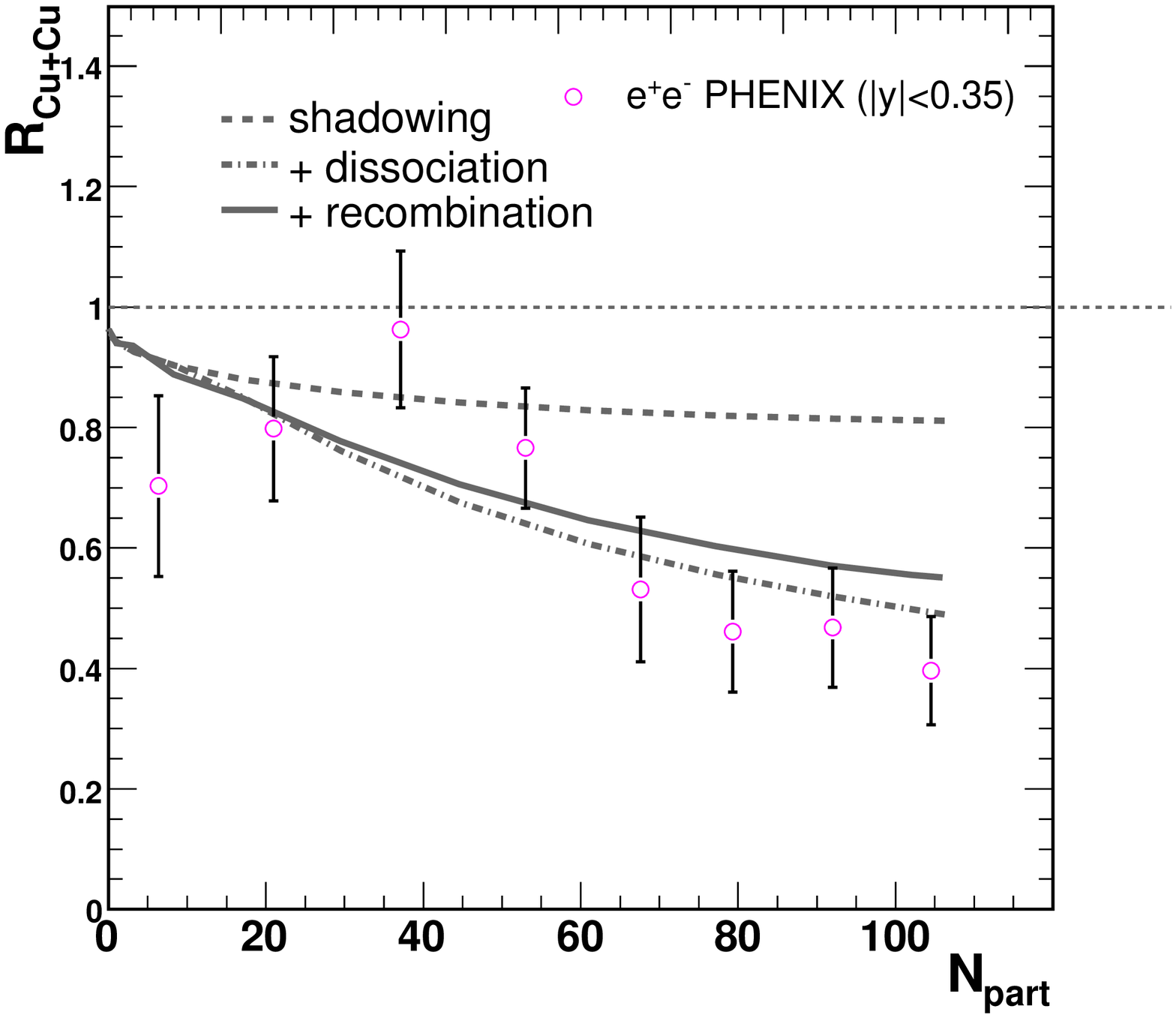}%
    \includegraphics[width=.5\linewidth]{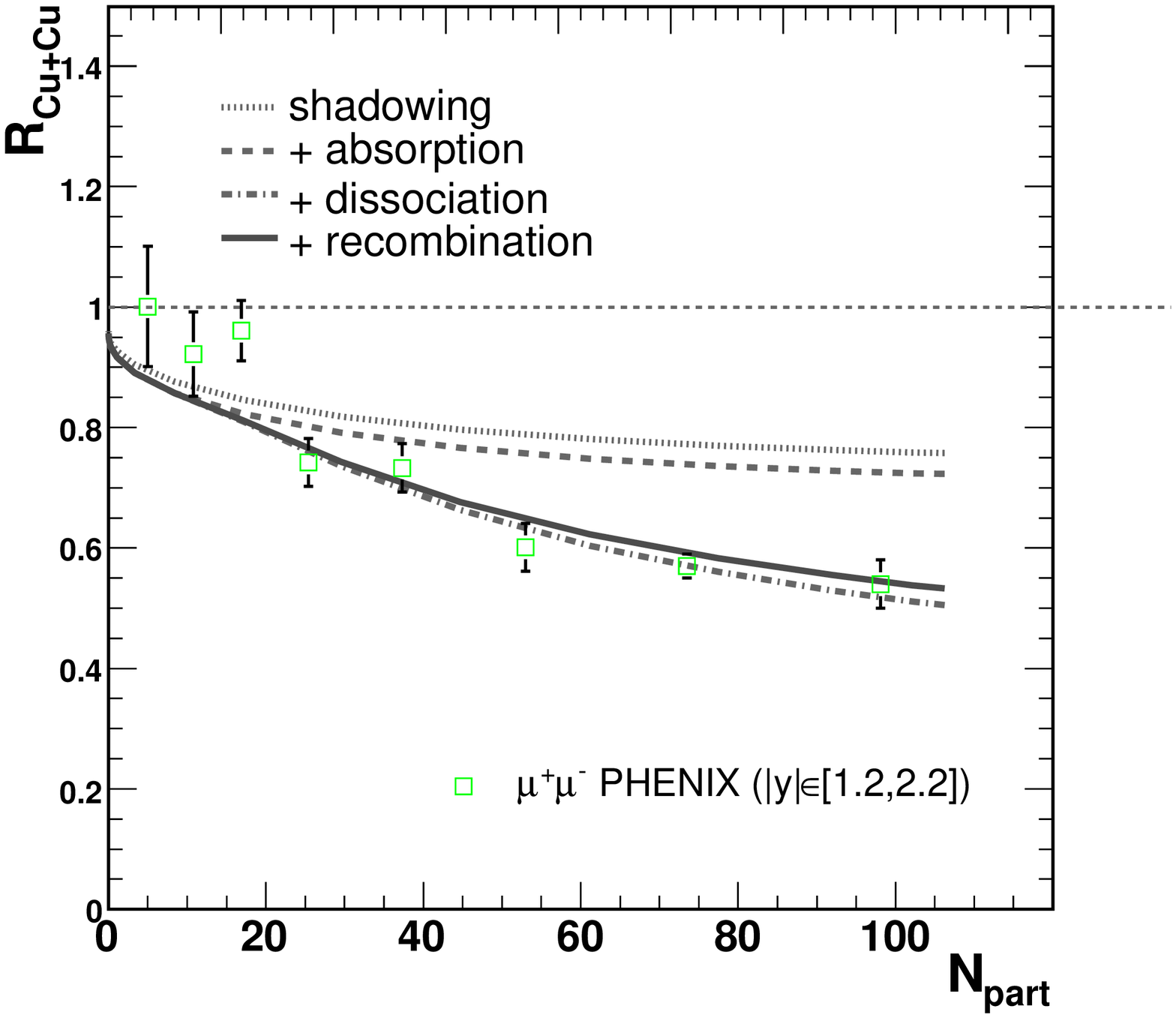}
  \end{center}
  \caption{Results for $J/\psi$ suppression in  {\it Cu+Cu} at RHIC
    ($\sqrt{s} = 200$ GeV), at mid- (left) and
    at forward rapidities (right). For details, see caption of
    Fig.~\ref{fig1}. Data are from
    \cite{Ci05}.}
  \label{fig2}
\end{figure}
For consistency, we have also made calculations for the $J/\psi$
suppression in {\it Cu+Cu} collisions at RHIC using the
same parameters as above for {\it Au+Au} collisions. The results are
shown in Fig.~\ref{fig2}, and are in good agreement with the 
experimental data, except maybe for peripheral collisions, where the 
error bars are quite large.
Concluding, our  procedure gives a reasonable description of data
both at mid- and forward rapidity for different collision systems at
RHIC. The effect of recombination is more pronounced at midrapidity.

\begin{figure}[t!]
  \begin{center}
    \includegraphics[width=.5\linewidth]{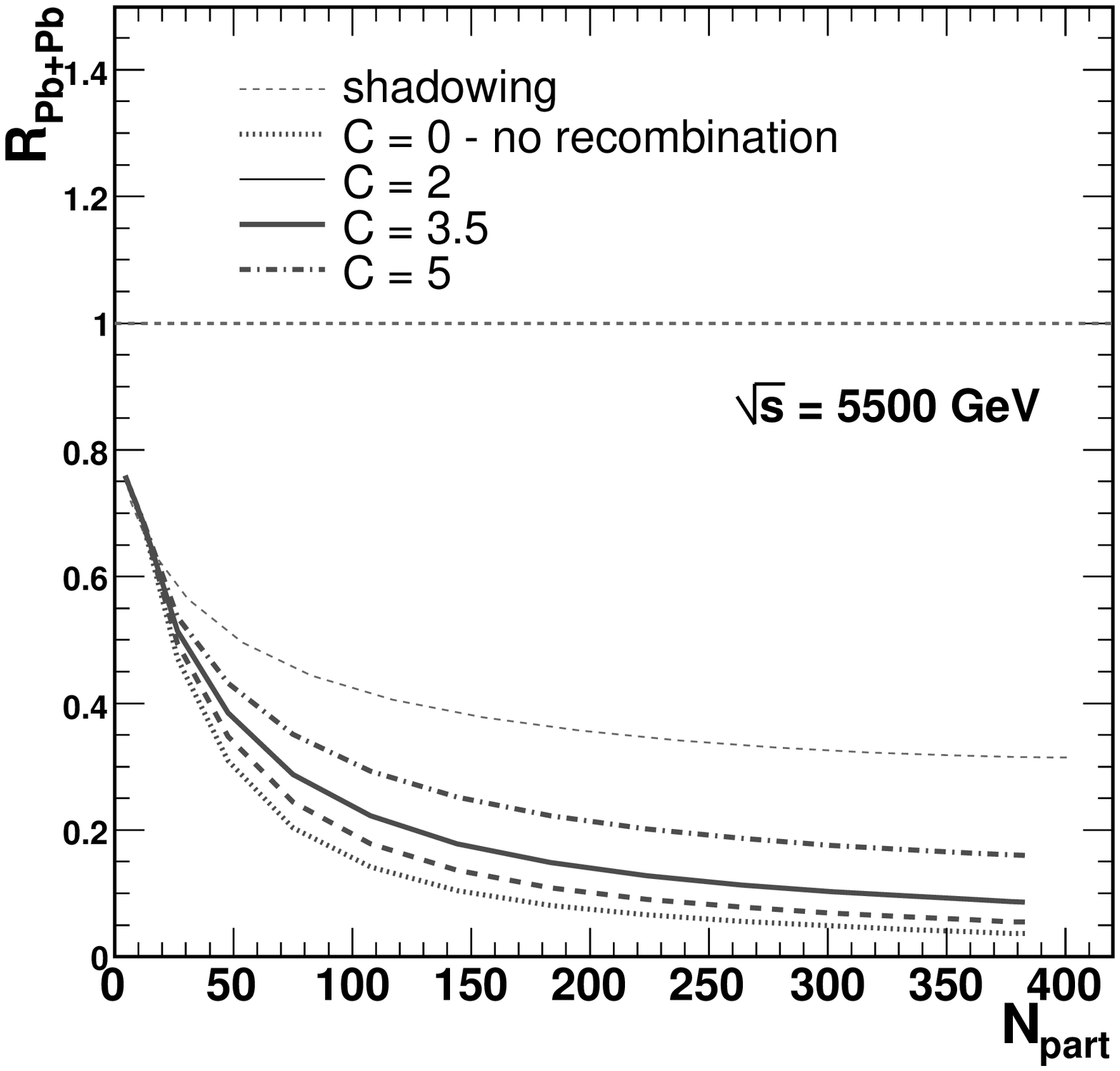}%
    \includegraphics[width=.5\linewidth]{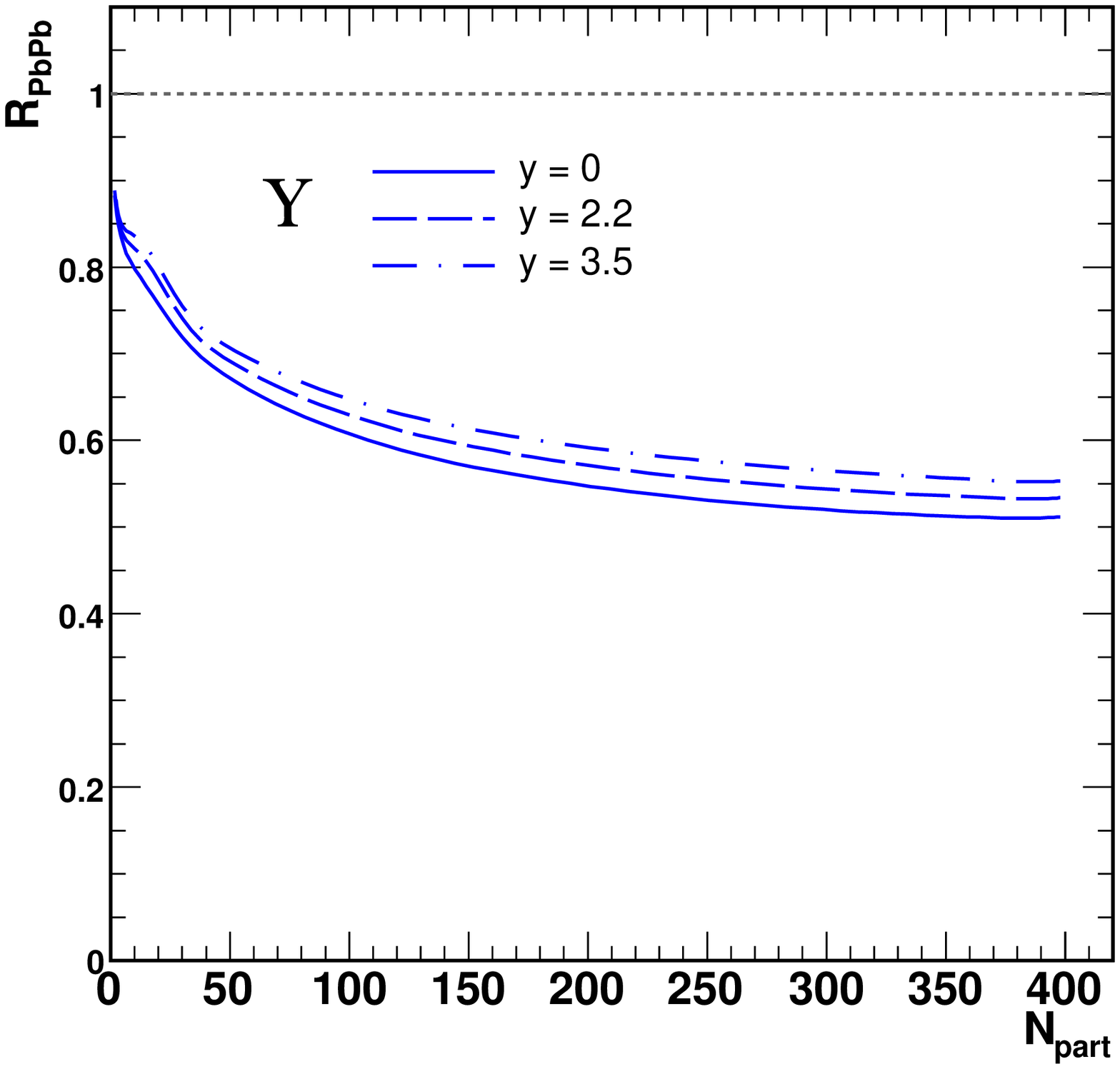}%
  \end{center}
\caption{Left: $J/\psi$ suppression in {\it Pb+Pb} at LHC 
($\sqrt{s} = 5.5$ TeV) at mid-rapidities for different values
of the parameter $C$. The upper line is the suppression due to 
initial-state effects (shadowing).
Right: Centrality dependence of $\Upsilon$ suppression due to 
gluon shadowing in {\it Pb+Pb} collisions at $\sqrt{s} = $ 5.5 TeV.} 
\label{fig3}
\end{figure}

Based on our previous discussion, it is obvious that 
dissociation-recombination
effects will be of crucial importance in {\it Pb+Pb} collisions at LHC
($\sqrt{s} = 5.5$ TeV). Assuming that the energy dependence of open 
charm and $J/\psi$ in {\it pp} collisions is the same (between RHIC 
and LHC energies), the energy dependence of the parameter $C$ will be 
that  of $\sigma^{\cc}_{pp}/\sigma_{pp}$. The total and differential 
cross section for charm can be calculated using perturbative 
techniques \cite{Cac05}. The calculations for low 
energies are in agreement with data, yet predictions for RHIC and 
Tevatron energies are lower than the data. Therefore, the 
extrapolation to LHC is quite uncertain. If we parameterize the 
energy dependence of open charm production as 
$\sigma^{\cc} \propto s^\alpha$, with $\alpha = 0.3$ and use the 
values of non-diffractive $\sigma_{pp}$ as 34\,mb for RHIC and
59\,mb for LHC, we obtain $C = 2.5$ at LHC -- a value about four
times larger than the corresponding one at RHIC. In view of that we
consider that realistic values of $C$ at LHC are of the range 2 to 3.
In Fig.~\ref{fig3}(a) we have calculated the $J/\psi$ suppression at 
LHC for several values of C, including the case of absence of
recombination effects ($C=0$).
Although the density of charm grows substantially from RHIC to LHC,
the combined effect of initial-state shadowing and comovers
dissociation appears to overcome the effect of parton recombination.
This is in sharp contrast with the findings of \cite{And07}, where
a strong enhancement of the $J/\psi$ yield with increasing centrality
was predicted. Note that in our approach only comovers and open charm 
produced at the same impact parameter as the initial $J/\psi$ are 
allowed to interact, whereas models assuming global equilibrium of 
the produced charm with the medium allow for recombination of $\cc$ 
pairs from the whole volume of the fireball.

Finally, we would like to discuss the impact of initial state effects 
on bottomonium production. The absorptive cross section for $\Upsilon$
is 40-50\% smaller than that for $J/\psi$ and $\psi'$, and 
energy-momentum conservation mechanisms are pushed to higher $x_F$ due 
to the large mass of the bottomonium. Therefore, nuclear absorption for 
$\Upsilon$ at LHC is expected to be quite small. The suppression of 
bottomonium due to gluon shadowing in Pb+Pb collisions at LHC is shown 
in Fig.~\ref{fig3}(b) for several rapidities. The 
suppression is about 50\% from mid-central to central collisions, and 
would be the same for all members of the $\Upsilon$ family. This 
establishes the baseline for further calculations
of bottomonium dissociation and recombination in the final state.

\section{Conclusions}
\label{concl}

The effects of recombination of $\cc$ pairs into $J/\psi$ are
incorporated in the comovers interaction model. These effects are 
negligible at low energies (SPS) due to the low density of open
charm. The model does not assume thermal equilibrium of the matter
produced in the collision and includes a comprehensive treatment of 
initial-state effects, such as shadowing, nuclear absorption and 
energy-momentum conservation.

In our approach, the magnitude of the recombination effect is 
controlled by the total charm cross section in ${\it pp}$ collisions. 
Using it as an input, the centrality and rapidity dependence of 
experimental data is reproduced both for {\it Au+Au} and {\it Cu+Cu} 
collisions at full RHIC energy.
For LHC we are lacking the experimental information and should rely 
on estimates from theoretical models. For a reasonable choice of 
parameters, we predict that the suppression observed at RHIC and 
lower energies will still dominate over the recombination effects.
This is  due to the large density of comovers and to the strong 
initial-state suppression at these ultra-relativistic energies.

\end{document}